\documentclass[10pt]{article}

\usepackage[english]{babel}

\usepackage[pdftex]{graphics,color}
\usepackage{newlfont,epsfig}

\begin{document}
\date{}
\begin{center}
{\Large\bf Atom-field entanglement in a bimodal cavity}
\end{center}
\begin{center}
{\normalsize G.L. De\c cordi and A. Vidiella-Barranco \footnote{vidiella@ifi.unicamp.br}}
\end{center}
\begin{center}
{\normalsize{ Instituto de F\'\i sica ``Gleb Wataghin'' - Universidade Estadual de Campinas}}\\
{\normalsize{ 13083-859   Campinas  SP  Brazil}}\\
\end{center}
\begin{abstract}
We investigate some aspects of the dynamics and entanglement of bipartite quantum system (atom-quantized field), 
coupled to a third ``external" subsystem (quantized field). We make use of the Raman coupled model; a three-level atom in a 
lambda configuration interacting with two modes of the quantized cavity field. We consider the far off resonance limit, 
which allows the derivation of an effective Hamiltonian of a two-level atom coupled to the fields. We also make a comparison
with the situation in which one of the modes is treated classically rather than prepared in a quantum field (coherent state).
\end{abstract}
\section{Introduction}
\label{intro}
Three interacting quantum systems may also be viewed as a bipartite system 
coupled to a third, ``external" system. The Raman coupled model, introduced some years ago 
\cite{eber90} constitutes an example of a simple, analytically solvable model 
involving three quantum subsystems; a three level atom coupled to two 
modes of the cavity quantized field. In the limit in which the excited atomic 
state is well far-off resonance, a simpler, effective two-level 
Hamiltonian may be derived, either by an adiabatic elimination of the upper atomic 
level \cite{eber90} or via a unitary transformation \cite{bose95}. In such a
procedure, energy (Stark) shifts arise and care must be taken, given that they may 
not always be neglected \cite{knig91}. In fact, the presence of the shifts
normally leads to a very different dynamics, e.g., from a non-periodic to a periodic 
atomic inversion. Several features of the dynamics of such model have already been  
investigated \cite{eber90}; in particular, quantum entanglement and possible applications 
to quantum computation \cite{agar04,garr04}. Here we are going to discuss a different aspect 
of that system; the influence of one of the fields (a cavity mode) on 
the dynamics of the atom as well as on the bipartite entanglement between the atom and the 
other cavity mode. We are going to consider different field preparations, such as coherent and 
thermal states, and we will also compare our results to the case in which one of the modes is not quantized, 
but treated as a classical field, instead. For simplicity, we do not take into account cavity losses or 
atomic spontaneous decay. Our paper is organized as follows: in Sec. II we present the model and solution. 
In Sec. III we discuss dynamical features with different preparations. In Sec. IV we present the 
evolution of bipartite entanglement, and in Sec. V we summarize our conclusions.

\section{The model and solution}
\subsection{Two quantized modes}

We consider a three level atom (levels 1,2,3) in interaction with two modes (mode 1, of frequency 
$\omega_1$ and mode 2, of frequency $\omega_2$) of the quantized field 
in a lambda configuration. Direct transitions between the lower levels 1 and 2 are forbidden. 
If the upper level (level 3) is highly detuned from the fields (detuning 
$\Delta$), the effective Hamiltonian may be written as \cite{bose95}
\begin{equation}
H_{eff}  =  \sigma_{11}E_{1}+\sigma_{22}E_{2}+\hbar\omega_{1}a_{1}^{\dagger}a_{1}+\hbar\omega_{2}a_{2}^{\dagger}a_{2}
 -\hbar\frac{g_{1}^{2}}{\Delta}\sigma_{11}a_{1}^{\dagger}a_{1}-\hbar\frac{g_{2}^{2}}{\Delta}\sigma_{22}a_{2}^{\dagger}a_{2}+\label{heffective}
 -\hbar\frac{g_{1}g_{2}}{\Delta}\left(a_{1}^{\dagger}a_{2}\sigma_{12}+a_{2}^{\dagger}a_{1}\sigma_{21}\right),
\end{equation}
where $\sigma_{12}$ and $\sigma_{21}$ are the transition operators between levels 1 and 2, 
$a_1(a_1^\dagger)$ is the annihilation (creation) operator of mode 1, 
$a_2(a_2^\dagger)$ is the annihilation (creation) operator of mode 2, and $g_1(g_2)$ are 
the couplings of the transitions $1-3$ and $2-3$, respectively. The effective
Hamiltonian is valid in the limit $g_1/\Delta(g_2/\Delta)\ll 1$, but under certain conditions, 
exact similar Hamiltonians may also be obtained \cite{wu96}. The Stark shift terms 
$-\hbar \frac{g_1^2}{\Delta}\sigma_{11}a_1^\dagger a_1$ and $-\hbar \frac{g_2^2}{\Delta}\sigma_{22}a_2^\dagger a_2$ 
are usually neglected \cite{eber90,garr04}, but one should be very careful, given that their inclusion results in a 
Rabi frequency depending linearly on the photon numbers $n_1$ and $n_2$. As a consequence, because of the shifts, 
the dynamics of the Raman coupled model becomes basically periodic, with Rabi frequency
\begin{equation}
\Omega_{n_{1},n_{2}}=\frac{\left[g_{1}^{2}n_{1}+g_{2}^{2}\left(n_{2}+1\right)\right]}{2\Delta}\label{rabifreq}
\end{equation} 
in contrast to the Rabi frequency, which is proportional to $\sqrt{n_1(n_2+1)}$, if the Stark shifts are 
neglected \cite{eber90}. 
Assuming an initial density operator of the product form,
\begin{eqnarray}
\rho\left(0\right)=\sum_{n_{1}n_{2}m_{1}m_{2}}^{\infty}\rho_{n_{1}n_{2}m_{1}m_{2}}\left|1;n_{1},n_{2}\right\rangle \left\langle 1;m_{1},m_{2}\right|,
\label{initialdo}
\end{eqnarray}
i.e., with the atom initially prepared in level 1, and the fields in generic states characterized by the coefficients 
$\rho_{n_{1}n_{2} m_{1}m_{2}}$. The full time-dependent density operator for the tripartite system may be written as

\begin{eqnarray}
\rho\left(t\right) && = \sum_{n_{1}n_{2}m_{1}m_{2}}^{\infty}\{ A_{n_{1}n_{2}m_{1}m_{2}}\left|1;n_{1},n_{2}\right\rangle \left\langle 1;m_{1},m_{2}\right|
 +  B_{n_{1}n_{2}m_{1}m_{2}}\left|2;n_{1}-1,n_{2}+1\right\rangle \left\langle 2;m_{1}-1,m_{2}+1\right|\nonumber \\
\nonumber\\&& +  C_{n_{1}n_{2}m_{1}m_{2}}\left|1;n_{1},n_{2}\right\rangle \left\langle 2;m_{1}-1,m_{2}+1\right|+h.c.\},\nonumber 
\label{densityop}\\
\end{eqnarray}
with coefficients
\begin{eqnarray}
A_{n_{1}n_{2}m_{1}m_{2}}&=&\rho_{n_{1}n_{2}m_{1}m_{2}}e^{i\nu_{n_{1}n_{2}m_{1}m_{2}}t}k_{1,n_{1},n_{2}}\, k_{1,m_{1},m_{2}}^{*},\nonumber \\
B_{n_{1}n_{2}m_{1}m_{2}}&=&\rho_{n_{1}n_{2}m_{1}m_{2}}e^{i\nu_{n_{1}n_{2}m_{1}m_{2}}t}k_{2,n_{1},n_{2}}\, k_{2,m_{1},m_{2}}^{*},\nonumber \\ 
C_{n_{1}n_{2}m_{1}m_{2}}&=&\rho_{n_{1}n_{2}m_{1}m_{2}}e^{i\nu_{n_{1}n_{2}m_{1}m_{2}}t}k_{1,n_{1},n_{2}}\, k_{2,m_{1},m_{2}}^{*},\nonumber
\end{eqnarray}
where
\begin{equation}
\nu_{n_{1}n_{2}m_{1}m_{2}} = \left(m_{1}-n_{1}\right)\left(\omega_{1}-\frac{g_{1}^{2}}{2\Delta}\right)+
\left(m_{2}-n_{2}\right)\left(\omega_{2}-\frac{g_{2}^{2}}{2\Delta}\right),\nonumber
\end{equation}
and

\begin{eqnarray}
k_{1,n_{1},n_{2}}\left(t\right) & = & \cos\left(\Omega_{n_{1},n_{2}}t\right) 
+ i\left[\frac{n_{1}-r^{2}\left(n_{2}+1\right)}{n_{1}+r^{2}\left(n_{2}+1\right)}\right]\sin\left(\Omega_{n_{1},n_{2}}\,t\right),\nonumber \\
k_{2,n_{1},n_{2}}\left(t\right) & = & \frac{2ri\sqrt{n_{1}\left(n_{2}+1\right)}}{\left[n_{1}+r^{2}\left(n_{2}+1\right)\right]}\sin\left(\Omega_{n_{1},n_{2}}\,t\right),\nonumber \label{coeff}
\end{eqnarray}
where the Rabi frequency $\Omega_{n_{1},n_{2}}$ is given in Eq. (\ref{rabifreq}) and having defined $r\equiv g_2/g_1$.

\subsubsection{Atomic dynamics with different initial field preparations}

\begin{figure}[ht]
% Use the relevant command for your figure-insertion program
% to insert the figure file.
% For example, with the option graphics use
\begin{center}
\includegraphics{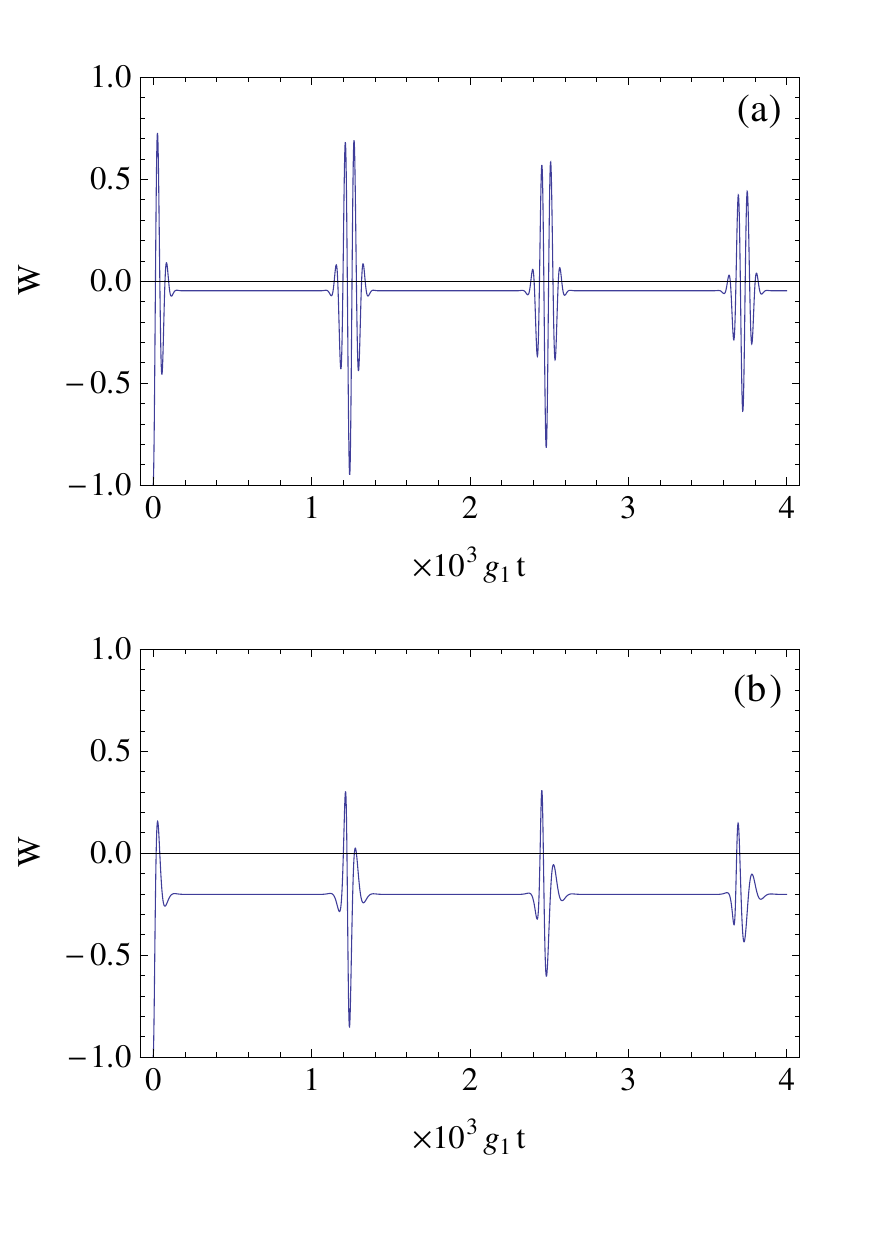}
\end{center}
% If not, use
%\vspace{5cm}       % Give the correct figure height in cm
\caption{Atomic inversion as a function of the scaled time $\tau = g_1 t$ for a) modes 1 and 2 
prepared in coherent states and b) mode 1 prepared in a 
coherent state and mode 2 prepared in a thermal state. 
In both cases $\overline{n}_1 = 10.5$ and $\overline{n}_2 = 10.1$.
We have considered $g_1 = g_2$ and $r = 1.012$.}
\label{figure1}       % Give a unique label
\end{figure}

\begin{figure}[ht]
% Use the relevant command for your figure-insertion program
% to insert the figure file.
% For example, with the option graphics use
\begin{center}
\includegraphics{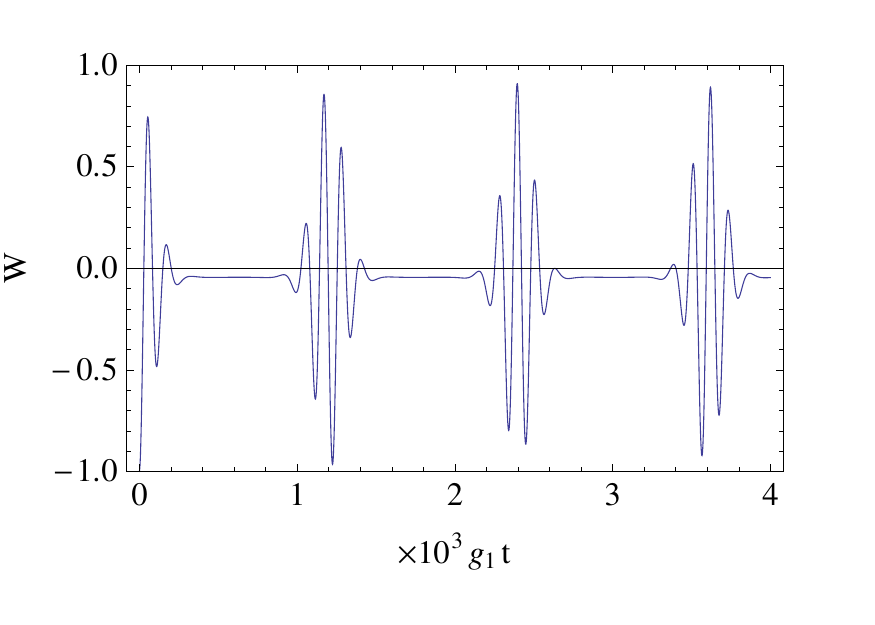}
\end{center}
% If not, use
%\vspace{5cm}       % Give the correct figure height in cm
\caption{Atomic inversion for mode 1 prepared in Fock state ($N = 5$) and mode 2 prepared in a coherent state ($\overline{n} = 5$), with $r = 1.023$.}
\label{figure2}       % Give a unique label
\end{figure}

The atomic response to the fields may be characterized by the atomic population inversion as a function of time, or
\begin{equation}
\label{atomicinv}
W\left(t\right)  =  2\, \mbox{Tr} \left[ \rho(t) |2\rangle\langle 2| \right] - 1 
 =  8\sum_{n_{1},n_{2}}^{\infty}p_{n_1} p_{n_2}
\frac{r^{2}n_{1}\left(n_{2}+1\right)}{\left[n_{1}+r^{2}\left(n_{2}+1\right)\right]^{2}}\sin^{2}\left(\Omega_{n_{1},n_{2}}t\right) - 1,\nonumber
\end{equation}
where $p_{n_1} p_{n_2}= \rho_{n_{1}n_{2}\,;\, n_{1}n_{2}}$ is the product of the photon number 
distributions of the initial fields. The atomic inversion has
peculiar features depending on the field statistics. For instance, in another well known model of optical 
resonance having a two level atom coupled to a single mode field, the Jaynes Cummings model (JCM), a field 
initially in a Fock state leads to pure oscillations of the atomic inversion, while an initial coherent state 
causes collapses and revivals \cite{eber80} of the Rabi oscillations.
On the other hand, in the JCM, for a field initially in a thermal state, the structure of collapses and revivals 
becomes highly desorganized \cite{radm82} and look like random. In the Raman coupled model, in which an effective two 
level atom is coupled to two fields, rather than one, we expect of course a different behaviour. Perhaps the most 
striking difference to the JCM is the periodicity of the atomic inversion; moreover, the ``revival'' 
times do not depend on the intensity of the fields, and the field statistics has only
little effect on the pattern of collapses and revivals \cite{knig91}. 
In order to illustrate this, in Fig. (\ref{figure1}) we have plots of the atomic inversion as a 
function of the scaled time $\tau = g_1 t$ for modes 1 and 2 prepared in coherent states: 
$p_{n_i} = e^{-\bar{n}_{i}}\bar{n}_{i}^{n_{i}}/n_{i}!$ [Fig. \ref{figure1}(a)] and for mode 1 prepared in a coherent state and mode 2 prepared in a 
thermal state: $p_{n_1}  = e^{-\bar{n}_{1}}\bar{n}_{1}^{n_{1}}/n_{1}!$ and $p_{n_2} = \bar{n}_{2}^{n_{2}}/\left(\bar{n}_{2}+1\right)^{n_{2}+1}$
[Fig. \ref{figure1}(b)]. 
We may note the periodical and well defined revivals occurring at the same times in both cases. Curioulsly, even in the 
case of a thermal (chaotic) initial preparation, revivals are well defined, although
their amplitude is slightly suppressed; probably an effect due to the broader thermal distribution. We also note that 
if one of the fields is prepared in a Fock state, if the other mode is prepared in a coherent state, for instance, collapses 
and revivals will still occur, due to the presence of the ``in phase" and ``out of phase 
terms" in Eq. (\ref{atomicinv}). In Fig. (\ref{figure2}) we have a plot of the atomic inversion for mode 1 prepared in a 
$N$ photons Fock state and mode 2 prepared in a coherent state, and we note again the pattern of periodic and regular 
collapses and revivals. In what follows, this particular preparation (Fock-coherent) is going to be compared with a situation 
termed ``partially classical", having one of the fields, mode 2, treated classically rather than prepared in a coherent state 
(fully quantized case).

\subsection{One quantized mode}

Now we consider mode 2 as being a classical field of amplitude $\Omega_L$;  the ``partially classical" case, keeping mode 1 quantized. 
The effetive Hamiltonian in the far-off resonance limit for the excited state is, 
in this case \cite{mato01},
\begin{equation}
H_{eff}^{\prime}=\frac{g^{2}a^{\dagger}a}{\Delta}\sigma_{11}+\frac{g^{2}r^{\prime}{}^2}{\Delta}\sigma_{22}+\lambda\left(\sigma_{12}a^{\dagger}
+\sigma_{21}a\right),\label{semclasshamil}
\end{equation}
where $g$ is the quantum field/atom coupling constant, $\lambda=g\left|\Omega_{L}^{*}\right|/\Delta$ 
is the effective coupling constant and $r^{\prime} = \left|\Omega_{L}^{*}\right|/g$. 
Note that apart from some shifts, the effective Hamiltonian is very similar to a JCM Hamiltonian. 
This means that the atomic response to a field (mode 1) prepared in a Fock state is going to be in the form of pure Rabi oscillations, 
in contrast to the case where mode 2 is a quantum field prepared in the ``quasi-classical" coherent state, which shows collapses and 
revivals. This is another example showing that regarding the atomic response to a field, an intense quantum coherent state is not 
equivalent to a classical field.

\section{Entanglement}

We would like now to discuss the entanglement between the atom and one mode of the field 
(mode 1, for instance) considering the other mode as an ``external" coupled
sub-system. We then trace over the variables of mode 2, initially prepared in a coherent state, 
and examine the degree of entanglement between the atom and the remaining 
field (mode 1) initially prepared, for the sake of simplicity, in a Fock state. In order to quantify 
entanglement, we use the negativity. As we have done in the case of 
the atomic inversion, we will compare the results with the case in which mode 2 is considered to be a 
classical field. For an initial state having the atom in level 1,
mode 1 in a Fock state $|N\rangle$ and mode 2 in a coherent state $|\alpha\rangle$, i.e., a product state 
$\left|\psi\left(0\right)\right\rangle =\left|1\right\rangle \otimes\left|N\right\rangle \otimes\left|\alpha\right\rangle$, 
the joint atom-mode 1 density operator, obtained after tracing over mode 2 from Eq. (\ref{densityop}), 
$\tilde{\rho}(t) = Tr_2[\rho(t)]$ is given by
\begin{eqnarray}
&&\tilde{\rho}\left(t\right)  =  \sum_{n=0}^{\infty}\left|c_{n}\right|^{2}k_{1,N,n}\, k_{1,N,n}^{*}\left|1;N\right\rangle \left\langle 1;N\right| 
+\sum_{n=0}^{\infty}\left|c_{n}\right|^{2}\, k_{2,N,n}\, k_{2,N,n}^{*}\left|2;N-1\right\rangle \left\langle 2;N-1\right|\nonumber \\
&&+F(t)\sum_{n=0}^{\infty}c_{n+1}c_{n}^{*}\, k_{1,N,n+1}\, k_{2,N,n}^{*}\left|1;N\right\rangle \left\langle 2;N-1\right|
+ h.c.,\nonumber \label{reduceddens}
\end{eqnarray}
with $F(t) = e^{-i\left(\omega_{2}-\frac{g_{2}^{2}}{2\Delta}\right)t}$ and $c_{n}=e^{-\bar{n}/2}\alpha^{n}/\sqrt{n!}$.

We may then calculate the negativity as a function of time

\begin{eqnarray}
{\cal {N}}\left(t\right) =2r\sqrt{\bar{n}N}\left\{ \Bigg[\sum_{n=0}^{\infty}e^{-\bar{n}}\frac{\bar{n}^{n}}{n!}\left[\frac{N-r^{2}\left(n+2\right)}{N+r^{2}\left(n+2\right)}\right]\right.
\frac{\sin\left(\Omega_{N,n}t\right)\sin\left(\Omega_{N,n+1}t\right)}{\left[N+r^{2}\left(n+1\right)\right]}\Bigg]^{2}\nonumber\\
\left.+\left[\sum_{n=0}^{\infty}e^{-\bar{n}}\frac{\bar{n}^{n}}{n!}\:\frac{\sin\left(\Omega_{N,n}t\right)\cos\left(\Omega_{N,n+1}t\right)}{\left[N+r^{2}\left(n+1\right)\right]}\right]^{2}\right\} ^{1/2},
\end{eqnarray}
where $\bar{n} = |\alpha|^2$ and $\Omega_{N,n}=\left[g_{1}^{2}N+g_{2}^{2}\left(n+1\right)\right]/2\Delta$. 
It is also worth to compare the negativity to the linear entropy relative to the atomic state (obtained after tracing over mode 1, $\rho_A(t) = 
Tr_1[\tilde{\rho}(t)]$), which is defined as $\zeta_A(t) = 1 - Tr[\rho_A^2(t)]$ or

\begin{equation}
\zeta_{A}\left(t\right)  =
2 \sum_{n=0}^{\infty}e^{-\bar{n}}\:\frac{\bar{n}^{n}}{n!}\frac{4r^{2}N\left(n+1\right)}{\left[N+r^{2}\left(n+1\right)\right]^{2}}\sin^{2}\left(\Omega_{N,n}t\right) 
-2\left\{ \sum_{n=0}^{\infty}e^{-\bar{n}}\:\frac{\bar{n}^{n}}{n!}\frac{4r^{2}N\left(n+1\right)}
{\left[N+r^{2}\left(n+1\right)\right]^{2}}\sin^{2}\left(\Omega_{N,n}t\right)\right\} ^{2}.\label{linearentro}
\end{equation}

\begin{figure}[ht]
% Use the relevant command for your figure-insertion program
% to insert the figure file.
% For example, with the option graphics use
\begin{center}
\includegraphics{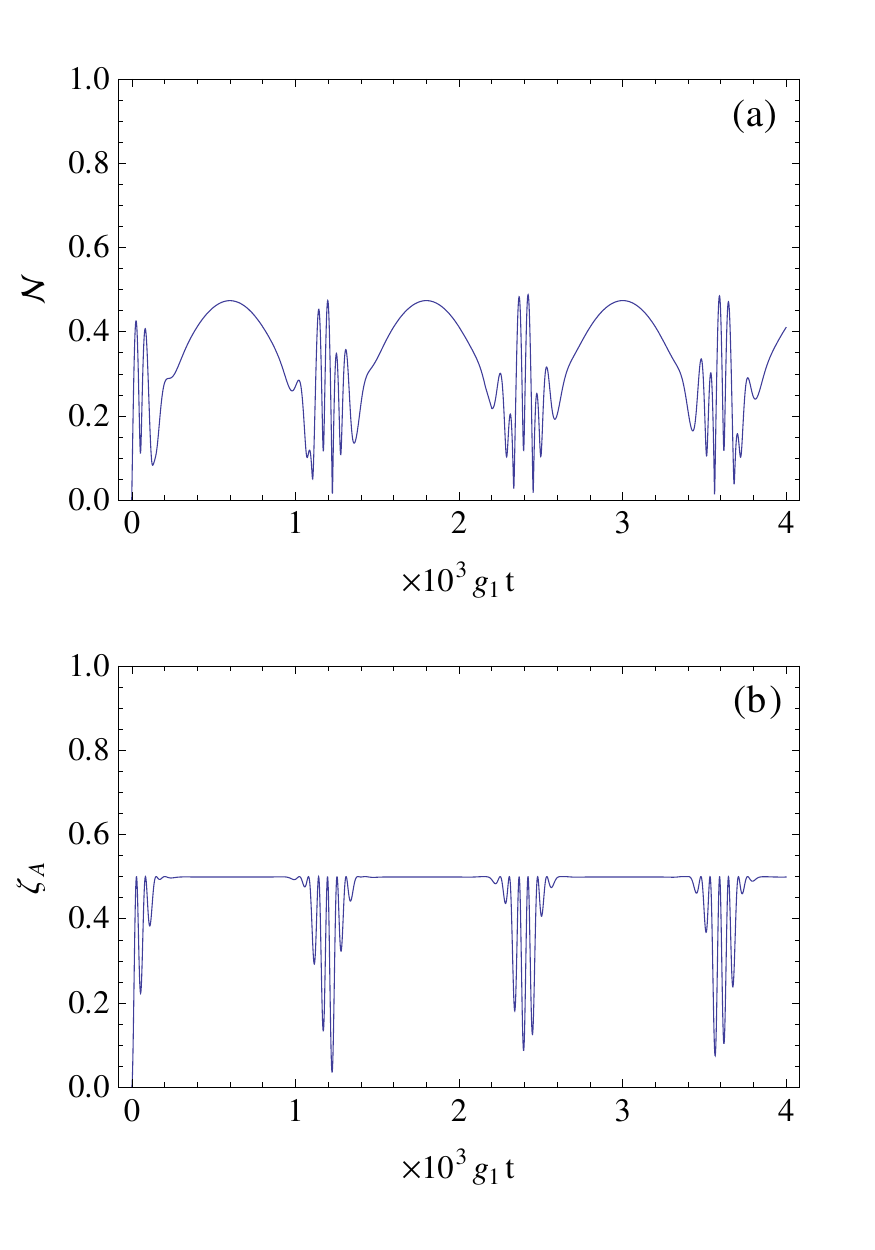}
\end{center}
% If not, use
%\vspace{5cm}       % Give the correct figure height in cm
\caption{Negativity (a) and linear entropy (b) as a function of the scaled time $g_1 t$
for mode 1 prepared in Fock state ($N = 5$) and mode 2 prepared in a coherent state ($\overline{n} = 5$), with $r = 1.023$.}
\label{figure3}       % Give a unique label
\end{figure}

\begin{figure}[ht]
% Use the relevant command for your figure-insertion program
% to insert the figure file.
% For example, with the option graphics use
\begin{center}
\includegraphics{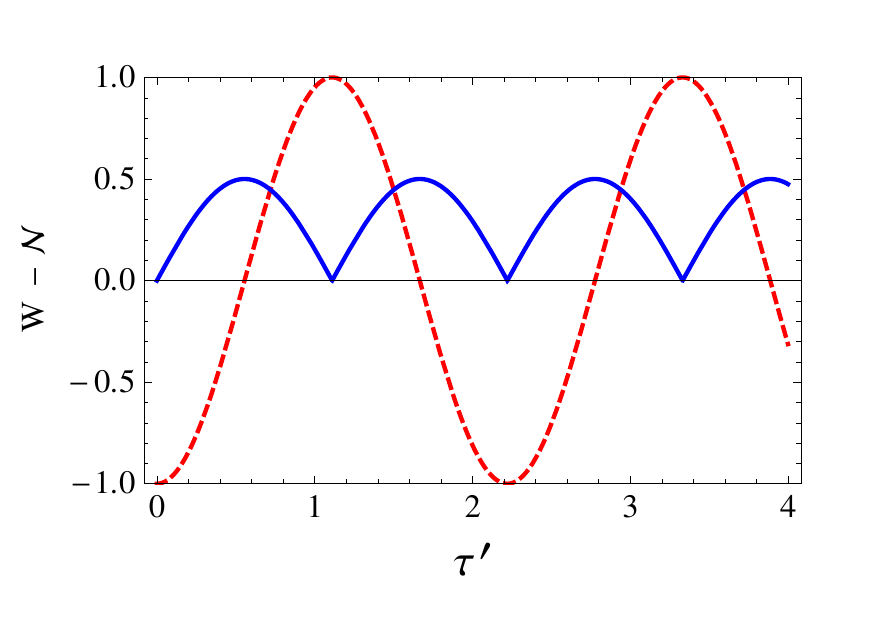}
\end{center}
% If not, use
%\vspace{5cm}       % Give the correct figure height in cm
\caption{Atomic inversion (red, dashed curve) and the Negativity (blue, continuous curve) 
as a function of the scaled time $\tau^{\prime}$ in the ``partially classical" case, having mode 1 prepared in a Fock state ($N = 2$) and $r^\prime = 1.41$.}
\label{figure4}       % Give a unique label
\end{figure}

The linear entropy may be used as a measure of the quantum state purity, as it it zero for a pure 
state and $> 0$ for a mixed state. In Fig. (\ref{figure3}) we have plotted the negativity (\ref{figure3}a) and 
the linear entropy (\ref{figure3}b) as a function of the scaled time $g_1 t$. 
We note that entanglement is maximum approximately in the middle of the collapse region, and the atom-mode 1 
state tends to become separable at the revival times themselves (see Fig. \ref{figure2}). At the same time one may notice the 
similarities and differences between the negativity and the linear entropy; the maximum of entanglement coincides
with the maximum of mixedness, and separability approximately occurs at times in which the atom is close to a pure state. 
However, we should point out that there are time intervals of maximum mixedness which do not correspond to maximum entanglement; 
this explicitly shows us the inadequacy of the linear entropy as a 
measure of entanglement (as expected), given that the state under consideration (atom-mode 1) is generally not a pure state. 
Again, the evolution of entanglement in the ``partially classical" case is going to be very different. We would like to remark 
that in this case a bipartite quantum system is under the action of an external classical field, instead of a  
tripartite system from which we have obtained a bipartite system by tracing over one of the subsystems. 
Having mode 1 prepared in a $N$ photon Fock state (mode 2 being a classical field), we have the following 
expressions for the atomic inversion (with the atom initially in state 1),
\begin{eqnarray}
W^{\prime}\left(t\right)=\frac{8r^{\prime2}N}{\left(N+r^{\prime2}\right)}\sin^{2}\left[\frac{\left(N+r^{\prime2}\right)}{2r^{\prime}}\tau^{\prime}\right]-1, \label{invsemiclass}
\end{eqnarray}
and the negativity,
\begin{equation}
{\cal {N^{\prime}}}\left(t\right) = \frac{2r^{\prime}\sqrt{N}}{\left(N+r^{\prime2}\right)}
\left\{\sin^{2}\left[\frac{\left(N+r^{\prime2}\right)}{2r^{\prime}}\tau^{\prime}\right]\right.
\left.-\frac{4r^{\prime2}N}{\left(N+r^{\prime2}\right)^{2}}\sin^{4}\left[\frac{\left(N+r^{\prime2}\right)}{2r^{\prime}}\tau^{\prime}\right]\right\}^{1/2},
\label{negatsemiclass} 
\end{equation}
where the scaled time is defined as $\tau^{\prime} = g\left|\Omega_{L}^{*}\right|t/\Delta$ and $r^{\prime} = \left|\Omega_{L}^{*}\right|/g$.
Compared to the fully quantized situation; there are no collapses and revivals, as the atomic inversion 
and the negativity are now periodic functions of time. This is illustrated in Fig. (\ref{figure4}), 
where we have plotted the atomic inversion in Eq. (\ref{invsemiclass}) and the negativity in Eq. (\ref{negatsemiclass}) 
as a function of the scaled time $\tau^{\prime}$. For a convenient choice of the parameter $r^{\prime}$, the atomic population 
completely inverts and returns to its initial state at times in which atom-mode 1 are in a separable state, 
a very different behaviour from what happens if mode 2 is a quantized field. 

\section{Conclusions}

We have presented a study of the dynamics of three coupled quantum systems: one three level atom and two quantized cavity 
fields, focusing on some properties of the atomic system (population inversion) and the atom-mode 1 bipartite system 
(entanglement). The second field (mode 2) has been basically traced out and treated as an ``external" 
subsystem. Our study has been based on the Raman coupled model, involving an effective two level atom coupled to two 
electromagnetic cavity fields. We have considered different preparations for mode 2: coherent and thermal states of the 
quantized field, as well as a classical field. In order to keep the consistency of the effective Hamiltonian, as has been 
already pointed out \cite{knig91,agar04} we have retained the Stark shift terms in the Raman Hamiltonian, which have the 
remarkable effect of keeping the dynamics periodic. Moreover, the periodic revival times will not depend on the statistics 
of the fields. This means that having one mode (or even two) prepared in the (highly mixed) thermal state will not change 
the regular atomic response during the atom/fields interaction, as we have shown explicitly here. This is a nice example of 
dynamical features robust against substantial variations in the field statistics.  We have addressed the issue of bipartite 
entanglement between the atom and mode 1, after tracing out mode 2: we have found that
for mode 1 initially prepared in a Fock state and mode 2 in a coherent state, the atom/mode 1 reach maximum entanglement 
in the collapse region of the atomic inversion. We have also compared both the atomic response and 
entanglement in the case in which mode 2 is treated as a classical field (``partially classical" case), rather than a 
quantum coherent ``quasi-classical" field, which result in very different evolutions.
\\
\begin{flushleft}
We thank Prof. C.J. Villas-B\^oas for fruitful comments.
This work is supported by the Agencies CNPq (INCT of Quantum Information) and FAPESP (CePOF), Brazil.
\end{flushleft}

%
% BibTeX users please use
% \bibliographystyle{}
% \bibliography{}

\begin{thebibliography}{}
%
% and use \bibitem to create references.
%
\bibitem{eber90} Christopher C. Gerry and J.H. Eberly, Phys. Rev. A \textbf{42}, (1990) 6805.

\bibitem{bose95} M. Alexanian and S.K. Bose, Phys. Rev. A \textbf{52}, (1995) 2218.

\bibitem{knig91} W.K. Lai, V. Bu{\v{z}}ek and P.L. Knight, Phys. Rev. A \textbf{44}, (1991) 6043.

\bibitem{agar04} Asoka Biswas and G.S. Agarwal, Phys. Rev. A \textbf{69}, (2004) 062306.

\bibitem{garr04}  J. Larson and B.M. Garraway, J. Mod. Opt. \textbf{51}, (2004) 1691.

\bibitem{wu96} Y. Wu, Phys. Rev. A \textbf{54}, (1996) 1586.

\bibitem{eber80} J.H. Eberly, N.B. Narozhny and J.J. Sanchez-Mondragon, Phys. Rev. Lett. \textbf{44}, (1980) 1323.

\bibitem{radm82} P.M. Radmore and P.L Knight, Phys. Lett. A \textbf{90}, (1982) 342.

\bibitem{mato01} M. Fran\c ca Santos, E. Solano, and R. L. de Matos Filho, Phys. Rev. Lett. \textbf{87}, (2001) 093601.
 
\end{thebibliography}
%
% Non-BibTeX users please use

%
% For one-column wide figures use
%

\end{document}